# ADS: Adaptive and Dynamic Scaling Mechanism for Multimedia Conferencing Services in the Cloud

Abbas Soltanian†, Diala Naboulsi†, Mohammad A. Salahuddin‡, Roch Glitho†, Halima Elbiaze‡, Constant Wette*
†Concordia University, ‡University of Waterloo, ‡Université du Québec À Montréal, *Ericsson, Canada
{ab_solta, d_naboul, glitho}@encs.concordia.ca, mohammad.salahuddin@ieee.org, elbiaze.halima@uqam.ca,
Constant.Wette.tchouati@ericsson.com

*Abstract*— Multimedia conferencing is used extensively in a wide range of applications, such as online games and distance learning. These applications need to efficiently scale the conference size as the number of participants fluctuates. Cloud is a technology that addresses the scalability issue. However, the proposed cloud-based solutions have several shortcomings in considering the future demand of applications while meeting both Quality of Service (QoS) requirements and efficiency in resource usage. In this paper, we propose an Adaptive and Dynamic Scaling mechanism (ADS) for multimedia conferencing services in the cloud. This mechanism enables scalable and elastic resource allocation with respect to the number of participants. ADS produces a cost efficient scaling schedule while considering the QoS requirements and the future demand of the conferencing service. We formulate the problem using Integer Linear Programming (ILP) and design a heuristic for it. Simulation results show that ADS mechanism elastically scales conferencing services. Moreover, the ADS heuristic is shown to outperform a greedy algorithm from a resource-efficiency perspective.

*Keywords*— Cloud Computing; Multimedia Conferencing; Platform-as-a-Service; Resource Allocation; Scaling Algorithm

## I. INTRODUCTION

Cloud computing is a paradigm in which resources (e.g., storage, network, and services) are provisioned rapidly on demand. It offers three main service models, including Software-as-a-Service (SaaS), Platform-as-a-Service (PaaS), and Infrastructure-as-a-Service (IaaS) [1]. Multimedia conferencing is the real-time exchange of media content (e.g., voice, video and text) between different parties [2]. It has several applications, such as massively multiplayer online games (MMOG) and distance learning. These applications have a considerable fluctuation in terms of the number of users. For instance, in one study, the number of players in the World of Warcraft game fluctuates between 1.5 and 2.5 million in 10 hours [3]. Therefore, such applications require a very high scalability and elasticity that cloud-based implementations may provide.

To speed up the provisioning of conferencing applications, providers can use conferencing services (e.g., dial-in video conferencing) available as SaaS [4], [5]. Therefore, the conferencing PaaS is responsible for ensuring that conferencing SaaSs deliver the required Quality of Service (QoS), such as delay or availability. A major challenge it faces is to ensure that the multimedia conferencing services offered as SaaS scale in an elastic manner, as participants join and leave. Required resources need to be available and cost efficiency needs to be ensured.

Fig. 1 depicts the assumed business model. It has four main roles as conferencing application providers, conferencing service providers, conferencing PaaS provider and, conferencing IaaS providers. Conferencing service providers use the conferencing PaaS to provision conferencing services and offer them as SaaS. Conferencing application providers also use the offered conferencing SaaSs to provision their applications. For instance, the provider of an online game application that allows dial-in audio conferencing between the game players uses Dial-in/Dial-out Audio conferencing service to provision the application. The conferencing PaaS provider offers the required functionalities for provisioning of the conferencing services (e.g., service hosting, execution, and management). Besides these functionalities, it provides a mechanism for scaling the conferencing services. The conferencing PaaS provider relies on the conferencing IaaS providers who offer the required resources (e.g., CPU and storage) to realize conferencing services. Also, there are several conferencing application users considered as conference participants.

During the conference, PaaS scales the conferencing services in response to the fluctuating number of participants. To ensure elastic scalability and to achieve cost efficiency, the conferencing PaaS requires an efficient scaling mechanism. This paper proposes an Adaptive and Dynamic Scaling mechanism (ADS) for conferencing services. ADS performs an elastic conference scaling, in terms of the number of participants, while meeting the Service Level Agreement (SLA) between the PaaS provider and the conferencing application providers. The dynamicity of ADS facilitates the on-demand scaling up or down of the conference. Moreover, its scaling policies change adaptively in accordance with the fluctuating number of conference participants to ensure elasticity.

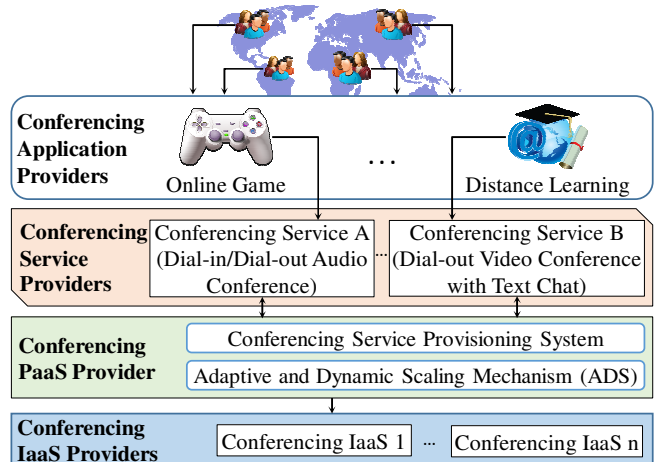

Fig. 1. Cloud-based Conferencing Business Model

We analyze the proposed scaling mechanism theoretically by modeling it as an optimization problem. Moreover, we design a heuristic for real-world scenarios. The results show the impact of ADS on the conference size. Also, we compare the cost pertaining to the resources and QoS of ADS with those of a greedy algorithm.

The rest of the paper is organized as follows. In Section II, we discuss the requirements of scaling the conferencing services as well as the related work. Section III delineates the system model, including cooperation and mathematical models. The ADS heuristic is presented in Section IV followed by the performance evaluation of ADS in Section V. We conclude in Section VI with contributions and future research directions.

## II. REQUIREMENTS AND RELATED WORK

In this section, we present the scaling requirements of multimedia conferencing services, followed by a review of the related work.

### A. Requirements

A crucial requirement of scaling mechanisms for multimedia conferencing services pertains to ***meeting the SLA*** between the conferencing application providers and the PaaS provider. For instance, each conferencing application can have a different acceptable delay for users before they can join the conference. Therefore, the scaling mechanism should ***utilize the application-level (high-level) metrics*** (e.g., the delay before users join the conference). Also, the decisions of the scaling mechanism should be aligned with the thresholds defined in the SLA.

Besides meeting the SLA, being ***cost efficient*** is another important requirement of scaling mechanisms. Inherently, resource provisioning with the minimum cost is desirable for any resource allocation system. In multimedia conferencing, since there is a significant fluctuation in the number of users, cost efficiency while scaling is of even greater importance. Moreover, ***elasticity*** is crucial in scaling. It increases resource utilization and reduces cost.

Another important requirement for scaling mechanisms of conferencing services is to ***consider the future demand*** of the conferencing applications. In fact, the scaling mechanism needs to be adaptive in accordance with the expected fluctuation in the future demands to ensure elasticity and avoid service outage.

### B. Related Work

The scaling mechanism presented in this work is used to solve the problem of efficient resource allocation for multimedia conferencing services. Since this problem entails the allocation of resources within the cloud environment, this is a cloud resource allocation problem. Based on the aforementioned requirements, we categorize the related work into 1) PaaS resource allocation solutions and, 2) IaaS resource allocation solutions.

*1) Existing PaaS Resource Allocation Solutions*

Several PaaS resource allocation solutions consider meeting the SLA requirements and cost efficiency. Anselmi *et al.* [6] model the resource allocation problem of PaaS as a Generalized Nash Equilibrium problem. Their scaling model relies on the number of virtual machines (VMs) that host the applications offered as SaaS. Satoh [7] also proposes a resource allocation mechanism for applications in the cloud. Satoh minimizes the required resources by reducing the redundant functions and data of the applications. Gomez *et al.* [8] introduce a PaaS framework that enables provisioning of cloud-based services and applications. Their platform also supports heterogeneous environments and relies on different IaaSs. Hu *et al.* [9] present an adaptive resource management algorithm. Their proposed PaaS dynamically allocates and reallocates the application instances based on the fluctuation of resource demands.

Although these works consider SLA and cost, they do not take the future demand of the applications into account. Moreover, their proposed solutions only consider low-level QoS requirements (e.g., demand on computing resources) and they do not utilize the application-level metrics in their solutions. Also, the scaling decisions in these solutions lead to adding a new instance of an application or adding a VM. Therefore, they support elasticity at the VM-level granularity.

Besides SLA and cost, there are some other works that consider future demand as well. Bunch *et al.* [10] present a pluggable auto-scaling mechanism for PaaS. They use an exponential smoothing algorithm to forecast the future demands. Roy *et al.* [11] also develop a model-predictive algorithm for workload forecasting. While [10] utilizes the application-level metrics, [11] does not and only considers low-level metrics for scaling. These solutions also can only support elasticity at the VM-level granularity.

*2) Existing IaaS Resource Allocation Solutions*

Besides PaaS resource allocation solutions, there are some other solutions proposed for IaaS. Xavier *et al.* in [12] and [13] propose resource allocation algorithms for audio and video services. The proposed solutions scale the resources at the VM-level while attempting to minimize the cost. They also consider meeting the user's quality of experience in their algorithms. However, they do not consider the future demands.

Negralo *et al.* [14] present algorithms for scaling resources by using load balancing and addition or removal of VMs. Scaling is triggered when a threshold value is reached for low-level metrics (e.g., CPU or bandwidth usage) defined beforehand. While the main focus of this work is cost efficiency, they do not consider SLA requirements. Moreover, they do not take the future demands into account. Unlike [14], Gong *et al.* [15] propose a scaling solution that considers future resource demands. They propose an elastic resource scaling mechanism to minimize the cost of resources. However, they do not consider the application-level metrics and meeting the SLA.

There are some other works, such as [16] and [17], with the objective of minimizing the wastage of resources. However, neither of them considers the future demand of the applications. Moreover, scaling decisions of these mechanisms result in addition or removal of a VM. Also, these solutions are not based on application-level metrics and they do not consider meeting the SLA.

TABLE I. Evaluation of the Related Works

| | State-of-the-art | SLA Awareness | Cost Efficiency | Elasticity | Future Demands | Application-level Metrics |
|---|---|---|---|---|---|---|
| **PaaS Solutions** | Anselmi et al. [1] | ✓ | ✓ | | | |
| | Satoh [2] | ✓ | ✓ | | | ✓ |
| | Gomez et al. [3] | ✓ | ✓ | ✓ | | |
| | Hu et al. [4] | ✓ | ✓ | | | |
| | Bunch et al. [5] | ✓ | ✓ | | ✓ | ✓ |
| | Roy et al. [6] | ✓ | ✓ | | ✓ | |
| | **ADS** | ✓ | ✓ | ✓ | ✓ | ✓ |
| **IaaS Solutions** | Xavier et al. [7] | ✓ | ✓ | ✓ | | ✓ |
| | Xavier et al. [8] | ✓ | ✓ | ✓ | | ✓ |
| | Negralo et al. [9] | | ✓ | | | |
| | Gong et al. [10] | ✓ | | ✓ | ✓ | |
| | Shen et al. [11] | | ✓ | | | |
| | Han et al. [12] | | ✓ | | | |

To conclude, all of the discussed related works try to allocate resources based on the existing or predicted number of users as the input. However, in this paper, we focus on deciding about the optimal size of the conferencing service while considering all requirements. In other words, ADS can complement the existing works. In fact, ADS can be used by all these solutions to tune their resource allocation mechanisms. For example, the derived conference size from ADS at each time slot can be used in [12] and [13] as the number of sources for their audio and video resource allocation algorithms. Moreover, these solutions can use the best scaling time derived from ADS as the trigger of their resource allocation. In Table I, the related works are evaluated with respect to the requirements of the scaling mechanisms for multimedia conferencing services.

### III. SYSTEM MODEL

Our system model includes cooperation and mathematical models. In our mathematical model, we define ADS as an Integer Linear Programming (ILP) problem.

*A. Cooperation Model*

According to the business model (c.f. Fig. 1), we consider a large-scale cloud environment to support scaling of the conferencing services. It consists of users as conference participants, a conferencing PaaS and, multiple conferencing IaaSs. The conference participants across a large geographical area want to join a conferencing application, such as MMOG. We assume there is an SLA between the conferencing application provider and the PaaS, where the QoS requirements are defined. One such requirement is the maximum acceptable delay for a participant to join the conference ($\theta$). Moreover, we assume there is an SLA between the conferencing PaaS and the conferencing IaaSs, where another set of QoS requirements are defined. One such QoS requirement is the time to provision resources in the IaaSs ($\delta$).

When a conference participant wants to join the conference, the required resources should be provisioned within $\theta$ time slots. In addition, when the scaling request is sent to the IaaSs, it takes $\delta$ time slots for resources to be provisioned. The challenge lies in finding the best time to send the scaling request. Moreover, this entails finding the required amount of resources to achieve the optimal resource cost while guaranteeing QoS requirements.

*B. Mathematical Model*

This subsection, presents ADS problem formulation which is modeled as an ILP problem.

*1) Problem Statement*

Given $n$ time slots of equal durations, let $A$ and $D$ represent the sets of expected arrivals and departures of conference participants, respectively. Such that, there will be a maximum of $a_i \in A$ and $d_i \in D$ participants, joining and leaving the conference during time slot $i$, respectively. It is assumed that $A$ and $D$ are available before the conference is started. Also, there is a threshold $\theta$ pertaining to the maximum acceptable delay before a participant can join the conference. We assume that $\theta$ is a multiple of time slots. Upon sending of the scaling request from the PaaS to the conferencing IaaSs, it is assumed that the required resources will be allocated within the time lag $\delta$. We assume that $\delta$ is a multiple of time slots and the $\delta$s for scaling up and scaling down are equal. Moreover, we assume the IaaS does not accept parallel scaling requests for the same conferencing service. Therefore, we assume there is at least $\delta$ time slots between two consecutive scaling requests. To simplify the problem, we consider the same $\delta$ for all IaaSs. In addition, we assume $\delta < \theta$. The goal is to find the optimal scaling schedule, such that the total amount of allocated resources in terms of the number of participants is minimized over the conference duration.

We model this as an ILP problem where we assume that each conference participant needs the same amount of resources to join the conference. Tables II and III delineate the inputs and variables of our problem, respectively.

TABLE II. Problem Inputs

| Input | Definition |
|---|---|
| $n$ | Total number of time slots in the entire conference duration |
| $A$ | A set of expected arrivals of conference participants, such that during time slot $i$, a maximum $a_i \in A$ participants join the conference, $1 \leq i \leq n$ |
| $D$ | A set of expected departures of conference participants, such that during time slot $i$, a maximum $d_i \in D$ participants leave the conference, $1 \leq i \leq n$ |
| $L$ | A set of number of conference participants, such that during time slot $i$, a maximum of $l_i \in L$ participants are in the conference for more than $\theta$ time slots, $1 \leq i \leq n$ |
| $\delta$ | The time lag, stipulated in the conferencing IaaS SLA for the response to the resource provisioning request. $\delta > 1$ time slot, otherwise the problem is trivial. |
| $\theta$ | Maximum acceptable delay for preparing the conference service |
| $M$ | A big enough constant |

TABLE III. Problem Variables

| Variable | Definition |
|---|---|
| $X$ | $n \times n$ matrix, where $x_{i,j}$ is the actual number of participants allocated to the service at time slot $i$ whose corresponding request is sent from PaaS to the IaaS at time slot $j$ |
| $Y$ | $n \times n$ matrix, where $y_{i,j}$ is the actual number of participants de-allocated from the service at time slot $i$ whose corresponding request is sent from PaaS to the IaaS at time slot $j$ |
| $R$ | A vector of binary variables, where $r_j = \begin{cases} 1, \text{if PaaS sends scaling request to IaaS at time slot } j \\ 0, \text{otherwise} \end{cases}$ |

*2) Objective*

We assume that the cost of using resources at each time slot depends on the total number of participants in the conference at that time slot. Our objective is to minimize the cost while considering other QoS requirements. We consider the provisioned resources in terms of the number of participants and the remaining time of the conference after provisioning the resources. The resource allocation and de-allocation for time slot $i$, for which the request is sent to IaaSs at time slot $j$ are represented as $x_{i,j}$ and $y_{i,j}$, respectively. Since the result of the scaling request will be ready after $\delta$ time slots, the remaining time of the conference after sending the scaling request at time slot $j$ will be $n - (j + \delta)$. Equation (1) depicts our objective.

$$minimize \left\{ \sum_{i=1}^{n} \sum_{j=1}^{n-\delta} (x_{i,j} - y_{i,j}) \times (n - (j + \delta)) \right\} \quad (1)$$

*3) Constraints*

To respect the maximum acceptable delay (i.e., threshold $\theta$), the allocated resources, in terms of conference participants, between time slot $i$ and $i + \theta$ should be greater than or equal to the expected number of participants arriving at time slot $i$. In other words, in the SLA between PaaS and the application providers, the conferencing PaaS guarantees that there will be no user waiting for more than $\theta$ time slots to be served before the conference ends. Equations (2) and (3) enforce this constraint. Note that the resources can be reserved before or after arrivals of users. It means that the scaling request time (i.e., $j$ in these equations) can be from the moment that conference was started until the end of the conference.

$$\sum_{j=1}^{i+\theta-\delta} x_{i,j} \geq a_i \quad \forall\, 1 \leq i \leq (n - \theta) \quad (2)$$

$$\sum_{j=1}^{n-\delta} x_{i,j} \geq a_i \quad \forall\, (n - \theta) < i \leq n \quad (3)$$

If there are some participants in the conference and PaaS provides them their required service, the conference size cannot be scaled down more than the number of participants who are remaining in the conference. In fact, the conference size cannot shrink before participants leave the conference, as in equations (4), (5) and (6).

$$\sum_{j=1}^{n-\delta} y_{i,j} \leq d_i \quad \forall\, 1 \leq i \leq \delta \quad (4)$$

$$\sum_{j=i-\delta}^{n-\delta} y_{i,j} \leq d_i \quad \forall\, \delta + 1 \leq i \leq n \quad (5)$$

$$\sum_{j=1}^{i-\delta-1} y_{i,j} = 0 \quad \forall\, \delta + 1 < i \leq n \quad (6)$$

The maximum amount of scaling down requests at each time slot cannot be more than the maximum of total allocated resources before that time slot. This is guaranteed in equation (7).

$$\sum_{i=1}^{n} \sum_{t=1}^{j} x_{i,t} \geq \sum_{i=1}^{n} \sum_{t=1}^{j} y_{i,t} \quad \forall\, 1 \leq j \leq n \quad (7)$$

Based on $A$ and $D$, the set $L$ can be defined, such that there will be a maximum of $l_i \in L$ participants in time slot $i$, who can be in the conference for more than $\theta$ time slots. Therefore, at each time slot, the prepared conference size should at least have the required resources for the participants who have been in the conference for more than $\theta$ time slots. Equation (8) represents this constraint.

$$\sum_{i=1}^{n} \sum_{t=1}^{j-\delta} x_{i,t} - \sum_{i=1}^{n} \sum_{t=1}^{j-\delta} y_{i,t} \geq l_j \quad \forall\, \delta < j \leq n \quad (8)$$

The conferencing IaaSs can accept the new scaling request from the PaaS after the previous request has been processed completely. Therefore, two consecutive scaling requests from the conferencing PaaS must be separated by $\delta$, as depicted in (9).

$$\sum_{j=i}^{i+\delta-1} r_j \leq 1 \quad \forall\, 1 \leq i \leq n - \delta \quad (9)$$

Moreover, any changes in the conference size made at time slot $j$, should be mapped to their scaling request at the same time slot as shown in equations (10) and (11). We assume $M$ is a big enough constant in these equations.

$$M \times r_j \geq x_{i,j} \quad \forall\, 1 \leq i, j \leq n \quad (10)$$

$$M \times r_j \geq y_{i,j} \quad \forall\, 1 \leq i, j \leq n \quad (11)$$

To avoid unnecessary resource allocation or de-allocation, there should be no scaling requests over the last $\delta$ time slots of the conference. In fact, such a request, if made, will take effect after the end of the conference. Through equation (12), we ensure that such requests are not sent.

$$r_j = 0 \quad \forall\, n - \delta < j \leq n \quad (12)$$

## IV. ADS HEURISTIC

Based on the proposed mathematical model, reaching the optimal solution for the large-scale scenarios is very time-consuming. Therefore, we propose an ADS heuristic as well to reach a sub-optimal solution in a reasonable time. The ADS heuristic tries to find the best schedule for scaling requests while respecting the SLAs. Algorithm 1 delineates the ADS heuristic. It iterates over the set of time slots throughout the conference. We consider the constants shown in Table II as the inputs of this algorithm. Also, the output of ADS algorithm is an integer array $S$ with $n$ elements. Each $s_i \in S$ represents the required scaling amount at time slot $i$. ADS heuristic has two main phases. In the first phase, it tries to find the minimum possible conference size and the best time for scaling the conference. In the second phase, it makes sure that all scaling requests are separated by at least $\delta$ time slots.

Since the cost depends on the amount of the provisioned resources and their usage over time, ADS heuristic is designed with the objective of reserving the least resources, as late as possible. The latest time should respect $\delta$ and $\theta$. Also, the minimum amount should respect the number of participants who

| **Algorithm 1.** ADS Heuristic |
|---|
| **Input:** |
| $n, \delta, \theta, A, D$; // same as the inputs of Table II |
| **Output:** $S$; // an schedule set of scaling decisions |
| 1. old_size ← 0 // previously provisioned size of the conference |
| 2. new_size ← 0 // conference size that should be provided for the future |
| 3. **For each** $i \in n$ **do** |
| 4.     min_size ← ∞ |
| 5.     best_t ← 0 |
| *Phase 1: Find the best possible time for sending the scaling request* |
| 6.     **For** t = $i + \delta \to i + \theta$ **do** |
| 7.        total_size ← 0 |
| 8.        **For** p=1→ t **do** |
| 9.           total_size ← total_size + $a_p - d_p$ |
| 10.        **end for** |
| 11.        **If** (min_size ≥ total_size) **Then** |
| 12.           min_size ← total_size |
| 13.           best_t ← t − $\delta$ |
| 14.        **end if** |
| 15.     **end for** |
| *Phase 2: Set the amount of scaling request for the best found time and move i to the next available time for sending request to the IaaSs* |
| 16.     new_size ← min_size |
| 17.     $S[best\_t]$ ← new_size − old_size |
| 18.     old_size ← new_size |
| 19.     $i$ ← best_t + $\delta$ − 1; // -1 because it is in the loop and $i$ for next cycle will be (best_t− $\delta$) |
| 20. **end for each** |
| **Return** $S$ |

are in the conference. Therefore, in phase 1, ADS tries to find the minimum size of the conference and the best time to send the scaling request. Based on the inputs, conference scaling takes $\delta$ time slots. Therefore, at each time slot $i$, ADS should consider the total conference size of $\delta$ time slots ahead. Also, new participants can wait up to $\theta$ time slots to join the conference. Thus, ADS can consider it as well and checks the total conference size up to $\theta$ time slots ahead. In consequence, since the objective is to find the minimum cost, ADS considers the minimum conference size between time slots $i + \delta$ and $i + \theta$.

In phase 2, ADS heuristic ensures that the consecutive scaling requests are separated by more than $\delta$ time slots. Moreover, it keeps track of the previous scaling request and its corresponding conference size. ADS compares the previous conference size with the result of phase 1 to decide about the scaling amount as the output of the algorithm. A positive value in the output means the request is to scale up, while a negative one means to scale down.

## V. SIMULATION RESULTS

In this section we will describe our evaluation scenarios and the simulation settings, followed by comparison results.

### A. Evaluation Scenarios and Simulation Settings

As the evaluation scenarios, we consider two different conferencing applications. (i) Massively Multiplayer Online Game (MMOG) and, (ii) Online Political Party Discussion (OPPD). In both scenarios, the users as the conference participants, are sharing their videos and audios in the logic of the application. In MMOG, users join and leave the game from all over the world. Thus, there is a significant fluctuation in the number of participants. In contrast, in OPPD, since the participants are limited, the fluctuation of the conference size is small.

For our simulation, we randomly generate the number of participants joining and leaving the conference at each time slot. To cover all possibilities, we keep the same conference size over a part of this time. This means that either no one joins or leaves the conference, or the number of users joining the conference is equal to the number of users leaving at each time slot, over that part. In our simulation, we divide the conference duration to 100 time slots. Also, we assume the resource provisioning time and the acceptable delay are 3 and 4 time slots, respectively. In addition, we set the fluctuation of the number of users to up to 1500 and 300 in MMOG and OPPD, respectively. Simulation parameters and settings are depicted in Table IV.

### B. Results

We implement the ADS algorithm in JAVA. Also, we use the LPSolve engine [18] to find the ADS optimal solution for our mathematical model. We compare the results of our algorithm with that of the optimal solution and the expected conference size. Also, we use a greedy algorithm as the baseline of our comparison. Since there is no similar heuristic in the literature that meets all of our requirements, this allows us to assess how our heuristic performs with respect to a simple greedy approach. The greedy algorithm operates on a periodic basis with a period equal to $\delta$. At time slot $t$ (with $t \bmod \delta = 0$), it derives the maximum number of participants between time slots $t + \delta$ and $t + 2\delta$. It then scales the conference accordingly. By that, the greedy approach is capable of satisfying the threshold of user's acceptable delay. Fig. 2 and 3, depict the created conference size for MMOG and OPPD applications, respectively. As these figures show, both our optimal and heuristic solutions can scale the conference size up and down. The scaling is elastic and it respects the SLAs.

Although in our scenarios, users can wait up to $\theta$ time slots to join the conference, there could be a cost for the delay as QoS violation. Fig. 4 and 5 show the total resource allocation and QoS violation costs of our scaling mechanism for MMOG and OPPD, respectively. As shown in these figures, the ADS heuristic outperforms the greedy algorithm from a resource-efficiency perspective. It leads to a solution that is closer to optimality with respect to the solution of the greedy algorithm, implying lower resource cost. However, this comes at the cost of a higher QoS violation. By comparing the solutions obtained from different algorithms, we notice that the greedy approach implies the least cost of QoS violation. It is followed by our ADS heuristic, while the ADS optimal solution leads to the highest QoS violation cost. These results highlight the trade-off that exists between the resource efficiency and QoS.

Fig. 4 and 5 also show that the cost of the ADS heuristic for provisioning resources in OPPD and MMOG has an 18% and a

TABLE IV. Simulation Parameters and Settings

| General Parameters | Value | MMOG Settings | | OPPD Settings | |
|---|---|---|---|---|---|
| $n$ | 100 | | | | |
| $\delta$ | 3 | $A$ and $D$ Fluctuation | 0-1500 | $A$ and $D$ Fluctuation | 0-300 |
| $\theta$ | 4 | | | | |
| $M$ | 1000000 | | | | |

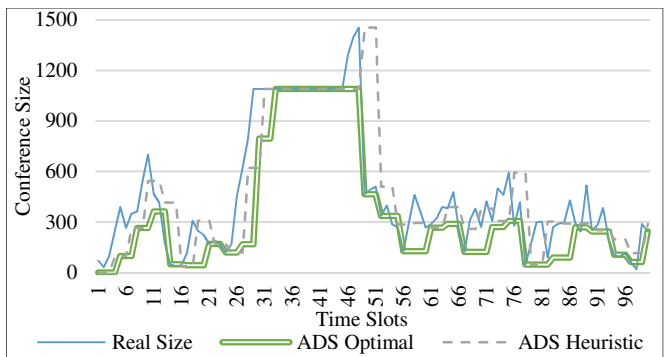

Fig. 2. Conference Size Comparison in MMOG

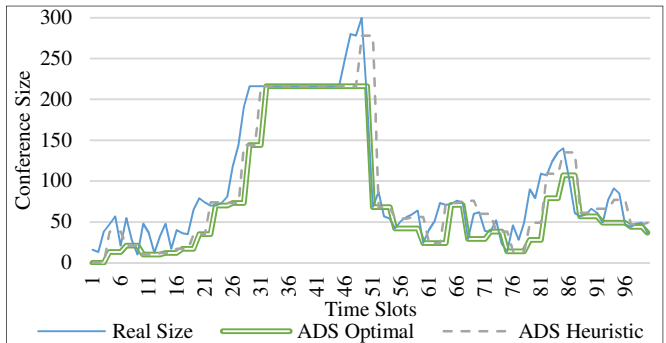

Fig. 3. Conference Size Comparison in OPPD

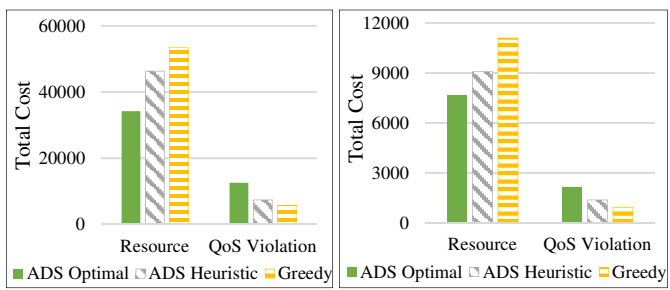

Fig. 4. Costs of Resources and QoS Violation in MMOG

Fig. 5. Costs of Resources and QoS Violation in OPPD

35% gap from the optimal solution, respectively. It means that the ADS heuristic can perform better when scaling conferences with lower fluctuations.

## VI. CONCLUSION

In this paper, we propose a novel Adaptive and Dynamic Scaling mechanism (ADS) for multimedia conferencing applications. ADS produces a cost-efficient scaling schedule while considering the QoS requirements and the future demands of the conferencing services. We model ADS as an optimization problem and design a heuristic to solve it in large-scale scenarios. Simulation results show the elasticity of ADS mechanism for conferencing services. Moreover, we show that the proposed ADS heuristic outperforms a simple greedy algorithm from a resource-efficiency perspective. In the future, we plan to extend our work by accounting for QoS violation cost, considering different resource allocation times from IaaSs, and having lower threshold of acceptable delay than the resource provisioning time.

## ACKNOWLEDGMENT

This work is supported in part by Ericsson and the National Science and Engineering Research Council (NSERC) of Canada.